\documentclass[preprint]{aastex61}
\usepackage{amsmath}
\usepackage{graphicx}
\usepackage[outdir=./]{epstopdf}
\usepackage{natbib}
\usepackage{multirow}
\shorttitle{Synchronization of the small scale}
\begin{document}
%%======================================
\title{Synchronization of small-scale magnetic features, blinkers, and coronal bright points}
\correspondingauthor{Hossein Safari }
\email{safari@znu.ac.ir}

\author{Zahra Shokri}
\affil{Department of Physics, Faculty of Science, University of Zanjan, University Blvd., 45371-38791, Zanjan, Iran}
\author{Nasibe Alipour}
\affil{Department of Physics, Faculty of Science, University of Zanjan, University Blvd., 45371-38791, Zanjan, Iran}
\author{Hossein Safari}
\affil{Department of Physics, Faculty of Science, University of Zanjan, University Blvd., 45371-38791, Zanjan, Iran}
\author{Pradeep Kayshap}
\affil{Vellore Institute of Technology (VIT), Bhopal University, Kothri Kalan, Sehore, M.P., India}
\author{Olena Podladchikova}
\affil{Physikalisch-Meteorologisches Observatorium Davos, World Radiation Center, 7260, Davos Dorf, Switzerland}
\author{Giuseppina Nigro} 
\affil{Department of Physics, University of Calabria, P. Bucci 31C, 87036 Rende (CS), Italy}.
\author{Durgesh Tripathi} 
\affil{Inter-University Centre for Astronomy and Astrophysics, Post Bag - 4, Ganeshkhind, Pune 411007, India}.
%%======================================

\begin{abstract} 
 We investigate the relationship between different transients such as blinkers detected in images taken at 304~{\AA}, extreme ultraviolet coronal bright points (ECBPs) at 193~{\AA}, X-ray coronal bright points (XCBPs) at 94~{\AA} on AIA, and magnetic features observed by HMI during ten years of solar cycle 24. An automatic identification method is applied to detect transients, and the YAFTA algorithm is used to extract the magnetic features. Using ten years of data, we detect in total 7,483,827 blinkers, 2,082,162  ECBPs, and 1,188,839 XCBPs, respectively, with their birthrate of about $1.1\times10^{-18}$ ${\rm m}^{-2}{\rm s}^{-1}$, $3.8\times10^{-19}$ ${\rm m}^{-2}{\rm s}^{-1}$, and $1.5\times10^{-19}$ ${\rm m}^{-2}{\rm s}^{-1}$. We find that about 80\% of blinkers are observed at the boundaries of supergranules, and 57\% (34\%) are associated with ECBPs (XCBPs). We further find that about 61{--}80\% of transients are associated with the isolated magnetic poles in the quiet Sun and that the normalized maximum intensities of the transients are correlated with photospheric magnetic flux of poles via a power law. These results conspicuously show that these transients have a magnetic origin and their synchronized behavior provides further clues towards the understanding of the coupling among the different layers of the solar atmosphere. Our study further reveals that the appearance of these transients is strongly anti-correlated with the sunspots cycle. This finding can be relevant for a better understanding of solar dynamo and magnetic structures at different scales during the solar cycle.

\end{abstract}
%========================================
\section{INTRODUCTION} \label{sec:intro}
%========================================
The inhomogeneous complex nature of the quiet Sun (QS) is predominantly due to the presence of a large variety of small-scale magnetic features. It is believed that these small-scale magnetic features are essentially the sources of magnetic energy, which can be dissipated to heat the upper atmosphere \citep{amari2015small}. Transients observed at different spatio-temporal scales, e.g., blinkers \citep{harrison1997euv, subramanian2012true}, explosive events \citep[e.g.,][]{GupT_2015}, coronal bright points \citep[CBPs,][]{vaiana1973x, golub1974solar, habbal1981spatial}, active region transient brightenings \citep[ARTBs,][]{berg2001}, Hi-C brightenings \citep[e.g.,][]{SubKT_2018, RajTK_2021}, transient loops in the core of active regions \citep{Tri_2021}, active region jets \citep[e.g.,][]{ChiYI_2008, MulTDM_2016}, etc. in the different atmospheric layers are associated with photospheric magnetic features. Due to such associations and various other properties of the transients, the process of magnetic reconnection is considered to be one of the key mechanisms for their generation. These transients observed ubiquitously throughout the solar atmospheric are further considered to play an essential role in transferring the mass and energy within the solar atmosphere \citep{parker1988nanoflares,schrijver1997sustaining,moore1999heating,winebarger2002energetics,yamauchi2005study,alipour2012automatic, priest2014book, UpenT_2021} and formation of the solar wind including switchbacks \citep{HasDL_1999,TuZM_2005,Zank_2020_theory,TriNS_2021,Liang_2021_Zankobs, UpenT_2022}.

Blinkers were first identified by \cite{harrison1997euv} and are defined as small-scale brightenings in the transition region. They are observed both in active regions (ARs) and QS, with AR blinkers being more abundant than those in QS \citep{parnell2002transition}. They were primarily identified in spectral lines such as \ion{O}{5}~629~{\AA}, \ion{O}{4}~554~{\AA}, \ion{O}{3}~599~{\AA}, \ion{He}{2}~304~{\AA}, and \ion{He}{1}~584~{\AA}, covering a temperature range of 5.0$\times 10^{4}$~K, to 2.5 $\times 10^{5}$ K \citep{harrison1997euv, berghmans1998quiet, harrison1999study}. Blinkers observed at 629~{\AA}, 554~{\AA}, and 599~{\AA} show a typical size of 20-50 Mm$^{2}$ with a lifetime 1-40 min \citep{harrison1997euv, bewsher2002transition}. \cite{berghmans1998quiet} showed that blinkers with sizes within the range of 40{--}400 Mm$^{2}$ are also recognizable at 304~{\AA} images with lifetimes about 2{--}60min.

X-rays and extreme ultraviolet (EUV) CBPs mostly contain small-scale loops \citep{ vaiana1973x, golub1974solar, habbal1981spatial}. Most of CBPs show similar morphological structures to large ARs with the small magnetic opposite polarity \citep{Madjarskacoronal}. Using the Skylab X-ray observations, \citet{golub1974solar} determined that 100 CBPs occur on the visible disk of the Sun at every time. Also, they found the lifetime of CBPs ranging from 2 to 48 hr. 
 \citet{alipour2015statistical} investigated the statistical properties of the CBPs in EUV emissions for 4.4 years. They showed that CBPs have a typical size of 130 Mm$^2$ and a lifetime ranging from a few minutes to more than three days. Also, they estimated the average number of 572 (ranging from 427 to 790) CBPs covering 2.6$\%$ solar visible disk at any time. Most (about 90\%) of CBPs  occur at the supergranular cell boundaries and junctions \citep{yousefzadeh2016motion}.
Recently, \citet{Hosseini2021} calculated a total energy-loss flux for the system of CBPs about $(4.84\pm1.60)\times 10^{3}$ erg cm$^{-2}$ s$^{-1}$. To estimate the contribution of QS small-scale events in the heating of the corona, they assumed a scale-free behavior for the energy of CBPs. Then, they obtained a total energy-loss flux  by extrapolating the power-law distribution to the nanoflare region (events with energies greater than $10^{24}$ erg). 
The energy-loss flux for the system of CBPs estimated 0.5{--}8.1\% of the quiet coronal plasma’s heat flux.

It is accepted that magnetic features play an important role in forming most events in the solar atmosphere. The magnetic fluxes observe in the form of bipole and unipole region at the solar surface \citep{2019Rubio}. A bi-pole refers to the two or a group of different opposite polarities that occur in a small region. Magnetic flux also occurs in the unipolar regions that are not distinctly coupled with any other pole(s) of opposite polarity in the vicinity. \citet{bewsher2002transition} showed that most blinkers occur above the regions of unipole or bipole with a strength of more than 10 G. Also, most of CBPs are associated with bipoles' magnetic fields \citep{Madjarskacoronal}. \citet{Sattarov2002ApJ} identified photospheric poles with strength above 20 G and sizes between 5.5{\arcsec} and 55.2\arcsec. They showed that the bipole numbers remained approximately constant (from 1992 to 2001). They also confirmed an anti-correlation of CBPs with sunspots.

There have been numerous studies on blinkers as well as CBPs on their physical and plasma properties. However, no work systematically studies the relationship between blinkers and CBPs observed in EUV and X-rays to the best of our knowledge. The relationship between transients and small-scale magnetic features is the main aim of this paper. Therefore, to identify, track and classify them, we use an automatic feature identification method based on the invariant and unique properties of the Zernike moments (ZMs) and the support vector machine (SVM) classifier. We have used this method essentially since the magnitude of ZMs is invariant  under rotation and image normalization \citep{alipour2012automatic,javaherian2014automatic,alipour2015statistical, Honarbakhsh2016, yousefzadeh2016motion, Raboonik2017, alipour2019prediction}. 

For the above-described purpose, we have extended the automatic identification method of \citet{alipour2015statistical} for the blinkers, ECBPs, and XCBPs by analyzing Atmospheric Imaging Assembly (AIA)/\textit{Solar Dynamics Observatory} (SDO) images in three channels (namely 304, 193, and 94 \AA). We aim to derive the statistical properties of blinkers, ECBPs, and XCBPs, as well as their relations. We also study the statistics of the photospheric magnetic features and their relationship with transients in the solar cycle 24.

This paper is organized as follows: a primary overview of data analysis and the automatic identification algorithms are explained in Section~\ref{sec:data and method}. The statistical analysis and results are presented in Section~\ref{sec:results}. Finally, we summarize and conclude in Section~\ref{sec:summery}.

%%-------------------------------------------------------------------
\section{Observations and methodology }\label{sec:data and method}
%%-------------------------------------------------------------------
\subsection{Data}
%%-------------------------------------------------------------------
The AIA on board SDO provides the solar atmospheric images at seven EUV and two ultraviolet (UV) wavelengths \citep{Lemen, Boerner}. AIA at 171, 193, and 211 \AA~ filters cover a temperature range 0.6 to 2 MK (plasma at QS temperature),  335, 94, and 131 \AA~ filters include temperature up to 10 MK (hot plasma), and 304 \AA~ filter is sensitive to plasma at 50000 K (most originate from the transition region). We used the full-disk synoptic images taken at \ion{He}{2}~ 304 \AA, \ion{Fe}{12}~ 193 \AA, and \ion{Fe}{18}~ 94 \AA~ filters. Coronal features such as ECBPs and XCBPs are detectable at 193 \AA~ EUV and  94 \AA~ X-ray emissions, respectively. However, we note that emissions at 94 \AA~ are contaminated signals from hot \ion{Fe}{18}~ line and cooler components of \ion{Fe}{10}~ and \ion{Fe}{14}~ \citep{Dwyer, zanna2011, aparna}. Also, blinkers as the transition region features are observable at 304 \AA~ emissions. The data set includes the sequence of co-spatiotemporal images with time intervals of 24 h. We analyzed the statistical properties related to transients for ten years from 2010 June until 2019 December. We obtained the co-spatiotemporal HMI continuum and line-of-sight (LOS) magnetogram images and applied a method to track solar photospheric flows on the intensity continuum images to determine the supergranular cell boundaries. 

We used the LOS magnetograms to study the statistics of magnetic patches and their relationship with blinkers, ECBPs, and XCBPs. To this end, we used the HMI-LOS magnetograms with the spatial sampling of 2.4 arcsec/pixel at 1024$\times$1024 pixels. The HMI instrument on board SDO provides different types of surface magnetic fields \citep{schou2012design, yeo2013,yeo_10gauss_bfielderror, couvidat2016observables}. The HMI noise levels depend on the HMI data product and location on the solar surface. We used more than 12 G and 20 G (i.e., HMI threshold values) to locate positive and negative polarities at the solar surface. These thresholds are higher than the maximum noise level of about 10 G for LOS magnetograms.

\subsection{Identification of AIA transients}\label{sec:method}
Identification methodologies of the solar features have been often employed based on the intensity characteristics at different observational wavelengths and determining a threshold value by the trial-and-errors \citep{bewsher2002transition, subramanian2012true}. Due to irregular variation of the intensity related to each feature (in different times, positions, and sizes), determining a single threshold seems to be accompanied by missing/false detection. Recently, automatic detection methods based on image moments and SVM to identify solar features have been developed \citep{alipour2012automatic,javaherian2014automatic,alipour2015statistical}. In this method, first, we extract information from the image moments covering the morphology, intensity, and geometry of features (blinkers, CBPs, etc.). Second, we fed the information to an SVM classifier to detect the objects. The method developed in this paper is a follow-up to the previous work presented by \citet{alipour2015statistical}, which identified the CBPs (mainly at 193 \AA) in EUV images taken by AIA. Here, we adopted the automatic identification method of \citet{alipour2015statistical} to recognize the blinkers at 304 \AA~ and XCBPs at 94 \AA~ filters in addition to identifying ECBPs at 193 {\AA}. Below we provide a summary of the automatic identification procedure applied here.

\begin{figure}
\centering
\vspace{1cm}
\includegraphics[width=\linewidth]{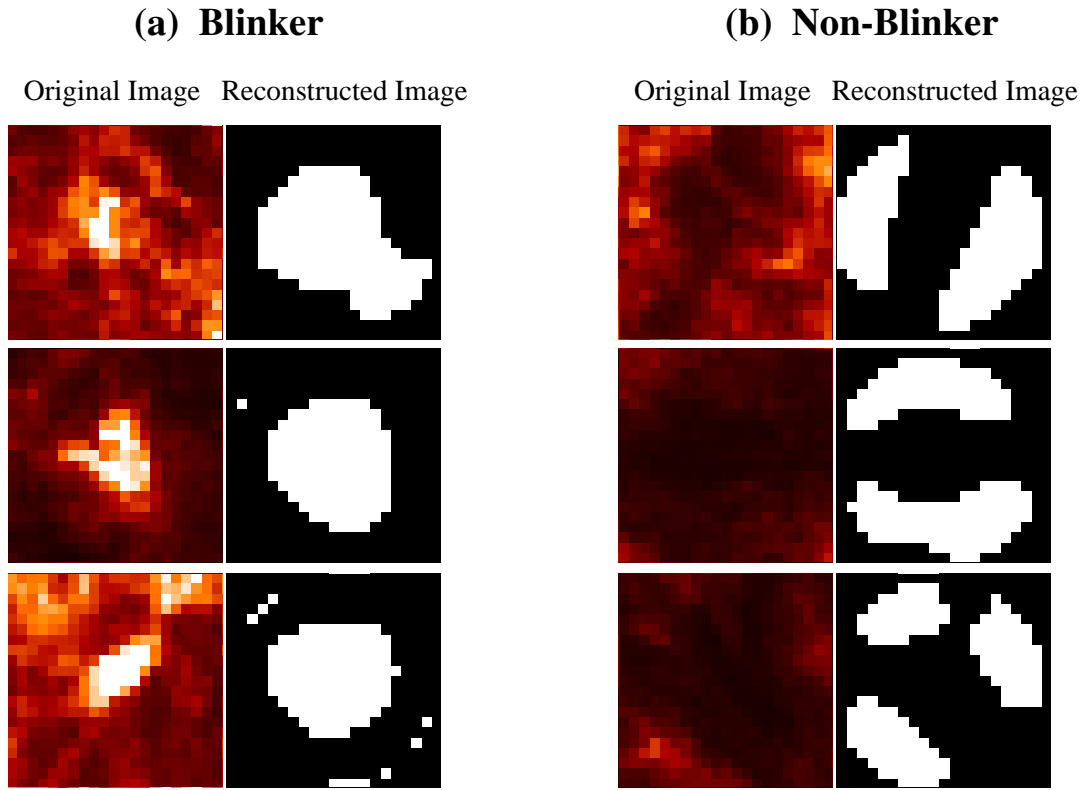}
\vspace{-1.5cm}
 \caption{ (a) Samples of blinkers observed by SDO/AIA at 304 \AA~ and reconstructed images with the maximum order number $p_{\rm max}$=5.  (b) Samples of the non-blinker regions and reconstructed images.}
\label{fig1}
\end{figure}

 \begin{itemize}
 \item[-] We performed a separate automatic identification classifier for each event, including event class (positive class) and non-event class (negative class).   
 
  \item[-] We collected (by visual inspection) a set of events by random surveying the data during ten years of observations. We selected the events based on well-known characteristics as (a) intensity criteria: localized intensity enhancement both temporally and spatially \citep{alipour2015statistical}, (b) size criteria: events with a length scale equal to or larger than 4.8{\arcsec} and smaller than 60\arcsec, and (c) morphology criteria: the small-scale loop-like or point-like structures. We selected about 1000 events according to the above criteria for each positive class.
  
 \item[-] We also collected some non-event in the negative class, including the regions without small-scale transient phenomena, parts of large-scale coronal loops, and regions outside of events. Figure~\ref{fig1} (left panel; a-b) shows the sample of original images related to blinkers and non-events at 304 \AA~ filter.

 \item[-] We computed the ZMs (with the order ($p$) and repetition number $(q)$) for each sub-image. The repetition number satisfies the condition that $|q|-p\leqslant 0$ is considered an even number. \citet{alipour2019prediction} showed that for an AR image, the ZMs with $ p_{\rm max} <10 $ reconstruct the overall shape of the AR. The morphological details of the AR were carefully reconstructed by increasing the order number up to 31. They also showed that for the $\rm p_{\rm max}(>31)$, the reconstructed images slightly deviated from the original one (see their Figure 1). We used $\rm p_{\rm max}=5 $, which is enough to reconstruct the overall shape of a  small-scale  event. Figure \ref{fig1} (right panel; a-b) shows the reconstructed images corresponding to blinkers and non-blinkers with different shapes and morphology, respectively.
 
\item[-] The algorithm automatically scans the full-disk images by a moving box in which the maximum intensity is located on the center of the box. SVM classifier picks up the events with their locations on the solar disk.   
 
 \end{itemize}
 
 \subsection{Performance of identification Method}
  To evaluate the performance of automatic identification for transients, we computed  various scores by applying the elements of a confusion matrix. The number of positive (P) and negative (N) cases in the data set are important terminologies for the confusions matrix \cite{powers2011}. The elements of the confusion matrix are true positive (TP: events correctly identified), false positive (FP: non-events incorrectly identified), true negative (TN: non-events correctly identified), and false negative (FN: events incorrectly identified) \citep{Fawcett}. The important scores are precision (positive and negative), recall (positive and negative), f$_{1}$ score (positive and negative), accuracy, Gilbert Skill Score (GS), Heidke skill score (HSS$_{1}$, HSS$_{2}$), and True Skill Statistic (TSS). Table \ref{tab1} gives the formulae for scores. For further details, refer to \citet{Barnes, Mason,Bloomfield, Bobra, Raboonik2017, alipour2019prediction}. The accuracy, precision, and f$_{1}$ are scores for class-balanced classifiers in which the number of events and non-events are approximately the same. Also, the HSS and GS scores suffer the class-imbalanced  problem. The TSS is an essential metric to measure the performance of the class-imbalanced  classifiers. 
  
  Here, we investigated the performance of three independent classifiers to identify blinkers, ECBPs, and XCBPs. To compute the skill scores as the measurement of performance of a classifier, we collected a database (by visual inspection) that includes more than 2000 events and 2000 non-events for each classifier. The database for each classifier contains features with various backgrounds, maximum intensity, shape, morphology, structures, and scales  collected within a solar cycle. The training data set includes about 70 percent (randomly selected) of the events (positive) and non-events (negative) classes. For the test data set, we used the remaining 30 percent of both positive and negative classes. We consider a sample event and non-event only in either the training or testing data set. We performed this randomized sampling for the training and test sets more than 100 times. We first calculated the elements of the TP, FP, TN, and FN for the test set of each trial. Then, using these elements, we computed the various skill scores. A similar analysis was performed to determine the performance of automatic detection for ECBPs and XCBPs. Table \ref{tab2} represents the average and standard deviation of the recall, precision, $f_{1}$ score (positive and negative), accuracy, Gilbert Skill Score (GS), Heidke Skill Score (HSS$_1$ and HSS$_2$), and True Skill Statistic (TSS) scores for automatic identification of blinkers, ECBPs, and XCBPs. We trained classifiers with various backgrounds, maximum intensities, shapes, morphologies, structures, and scales. Therefore, we expect the present identification method for the transients is less affected by the background variations. Some of the previous identification methods \citep{Sattarov2002ApJ, hara2003variation} for the transient events applied the intensity thresholds that highly depend on the background variation during a solar cycle. Applying this identification method to the SoHO/EIT image, \cite{alipour2015statistical} detected 670 CBPs compared to 450 events reported by \cite{Sattarov2002ApJ} at the same data set.

   \begin{table}
 \renewcommand{\arraystretch}{1.6}
\vspace{0.5cm}
\caption{ Definition of different skill scores. } \label{tab1} \centering
 \begin{tabular}{p{3.6cm}@{\hspace{15mm}}c}
		\hline
		Score & Formula \\ \hline \hline
		Recall (positive and negative) & $\rm{recall}^{+} = \dfrac{{\rm{TP}}}{\rm{TP} + \rm{FN}}$ \\ & ${\rm{recall}}^{-} = \dfrac{\rm{TN}}{\rm{TN} + \rm{FP}}$\\[2mm]\hline
		
		Precision (positive and negative) & ${\rm{precision}^{+}} = \dfrac{\rm{TP}}{\rm{TP} + \rm{FP}}$\\& ${\rm{precision}^{-}} = \dfrac{\rm{TN}}{\rm{TN} + \rm{FN}}$\\[2mm]\hline

		$f_{1}$ score (positive and negative) & $f_{1}^{+} = \dfrac{2 \times \rm{precision}^{+} \times \rm{recall}^{+} }{\rm{precision}^{+} + \rm{recall}^{+}}$\\ & $f_{1}^{-} = \dfrac{2 \times \rm{precision}^{-} \times \rm{recall}^{-} }{\rm{precision}^{-} + \rm{recall}^{-}}$\\[2mm]\hline

		Accuracy & $\rm{accuracy} = \dfrac{\rm{TP} + \rm{TN}}{\rm{TP} + \rm{FN} + \rm{TN} + \rm{FP}}$ \\[2mm]\hline

		Heidke Skill Score (HSS) & $\rm{HSS}_{1} = \dfrac{\rm{TP} - \rm{FP}}{\rm{TP} + \rm{FN}}$ \\[3mm] & $\rm{HSS}_{2} = \dfrac{2 \times [(\rm{TP}  \times \rm{TN}) - (\rm{FN}  \times \rm{FP})]}{(\rm{TP} + \rm{FN}) \times (\rm{FN} + \rm{TN}) + (\rm{TN} + \rm{FP}) \times (\rm{TP} + \rm{FP})}$ \\[2mm]\hline

		Gilbert Skill Score (GS)& $\rm{GS} = \dfrac{\rm{TP} - \rm{CH}}{\rm{TP} + \rm{FP} + \rm{FN} - \rm{CH}},$ \\[3mm] & $\rm{CH} = \dfrac{(\rm{TP} + \rm{FP}) \times (\rm{TP} + \rm{FN})}{\rm{TP} + \rm{FN} + \rm{TN} + \rm{FP}}$\\[2mm]\hline
		
		True Skill Statistic (TSS)&$\rm{TSS} = \dfrac{{\rm{TP}}}{\rm{TP} + \rm{FN}} - \dfrac{{\rm{FP}}}{\rm{FP} + \rm{TN}}$\\[2mm]\hline
	\end{tabular}
 \end{table}
 
\begin{table}
\vspace{0.5cm}
\caption{ Average and standard deviation of the precision (Precision$^{+}$ and Precision$^{-}$),
recall (Recall$^{+}$ and Recall$^{-}$), $f_{1}$ score (positive and negative), accuracy, Gilbert Skill Score (GS), Heidke Skill Score (HSS$_1$ and 
  HSS$_2$), and True Skill Statistic (TSS) scores  for automatic identification of blinkers, ECBPs, and XCBPs.}\label{tab2}
  \vspace{0.25cm}
\centering
	\begin{tabular}{cp{5cm}p{2cm}p{2cm}p{2cm}c}
		\cline{1-6}
        \vspace*{-5.5mm}
        \multirow{14}{*} {\rotatebox[origin=c]{90}{Scores}}\\
	     Features && Blinkers & ECBPs & XCBPs \\ \hline\hline
	   $ $ & $\rm{Recall}^{+}$ & $0.91\pm0.01$ & $0.91\pm0.01$ & $0.96\pm0.01$ & \\ \cline{2-6}
		$ $ &$\rm{Recall}^{-}$ & $0.97\pm0.01$ & $0.97\pm0.01$ & $0.89\pm0.01$ &\\ \cline{2-6}
		$ $ & $\rm{Precision}^{+}$ & $0.96\pm0.01$ & $0.97\pm0.01$ & $0.90\pm0.01$ & \\ \cline{2-6}
		$ $ & $\rm{Precision}^{-}$ & $0.91\pm0.01$ & $0.92\pm0.01$ & $0.96\pm0.01$ & \\ \cline{2-6}
		$ $ & $f_{1}$ score positive & $0.93\pm0.01$ & $0.94\pm0.01$ & $0.93\pm0.01$ & \\ \cline{2-6}
		$ $ & $f_{1}$ score negative & $0.94\pm0.01$ & $0.94\pm0.01$ & $0.93\pm0.01$ & \\ \cline{2-6}
		$ $ & $\rm{Accuracy}$ & $0.93\pm0.01$ & $0.94\pm0.01$ & $0.93\pm0.01$ & \\ \cline{2-6}
		$ $ & Heidke Skill Score $(\rm{HSS}_{1})$ & $0.87\pm0.01$ & $0.88\pm0.01$ & $0.86\pm0.01$ & \\ \cline{2-6}
		$ $ & Heidke Skill Score ($\rm{HSS}_{2}$) & $0.87\pm0.01$ & $0.88\pm0.01$ & $0.86\pm0.01$ & \\ \cline{2-6}
		$ $ & Gilbert Skill Score (GS) & $0.77\pm0.01$ & $0.79\pm0.02$ & $0.76\pm0.02$ & \\ \cline{2-6}
	    $ $ & True Skill Statistic (TSS) & $0.87\pm0.01$ & $0.88\pm0.01$ & $0.86\pm0.01$ & \\ \hline
	\end{tabular}
\end{table} 
 
 \subsection{Identification of magnetic patches from HMI magnetograms}  
\begin{figure}
\vspace{1cm}
\centering
\hspace*{-2.5cm}
\vspace*{-1.0cm}
\resizebox{10in}{!}{ \includegraphics{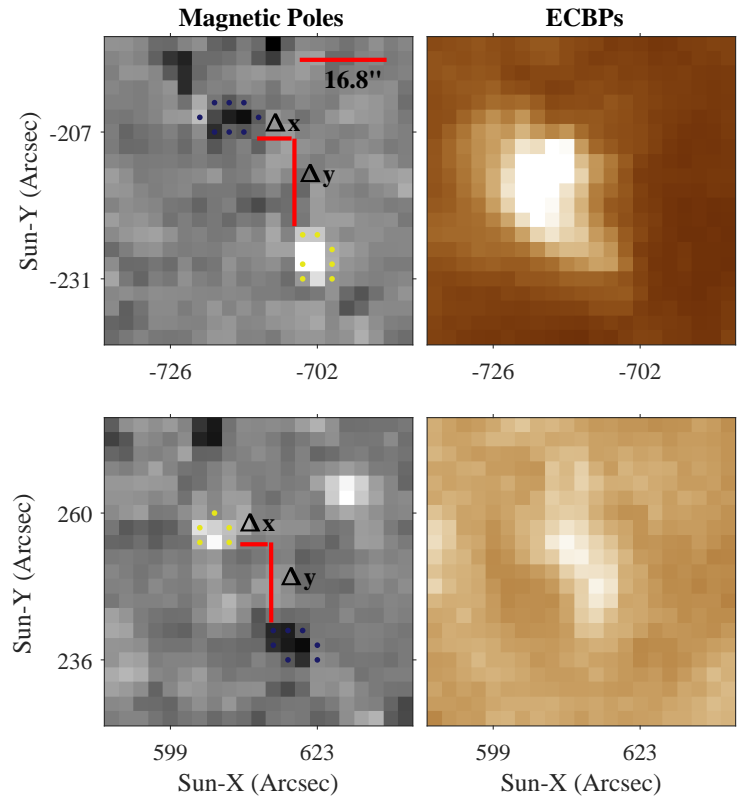}} 
 \caption{ Two samples of ECBPs and corresponding magnetograms. The red line indicates the distance between the opposite polarities. The scale line about 16.8{\arcsec} presents at the top of the magnetogram. }
\label{fig2}
\end{figure}

 Most transients are associated with unipolar or bipolar magnetic fields. So, we analyzed the statistics of the magnetic sources related to the features. To extract the small-scale magnetic poles, we used the Yet Another Feature Tracking Algorithm (YAFTA) developed based on the downhill method \citep{deforest2007ApJ}. We selected the negative and positive poles above 12 and 20 G with a length scale of about 4.8{\arcsec} to 60\arcsec. We extracted two nearest opposite polarities  with a separation less than 16.8{\arcsec} as a bipolar magnetic feature. To this end, we analyzed about 100 randomly selected ECBPs and their related magnetic bipoles. For most cases, we observe that the two nearest opposite polarity poles with a separation distance (from the boundary to boundary along with both the Sun-x and Sun-y axis) are less than 16.8{\arcsec} (for both $\Delta x$ and $\Delta y$) form a bipolar magnetic feature (Figure \ref{fig2}).
 
 \section{Results and Discussions}\label{sec:results}
\subsection{Statistics of transients}
\begin{figure}
\centering
\hspace*{+1.1cm}
\vspace*{-2.15cm}
\includegraphics[width=\linewidth]{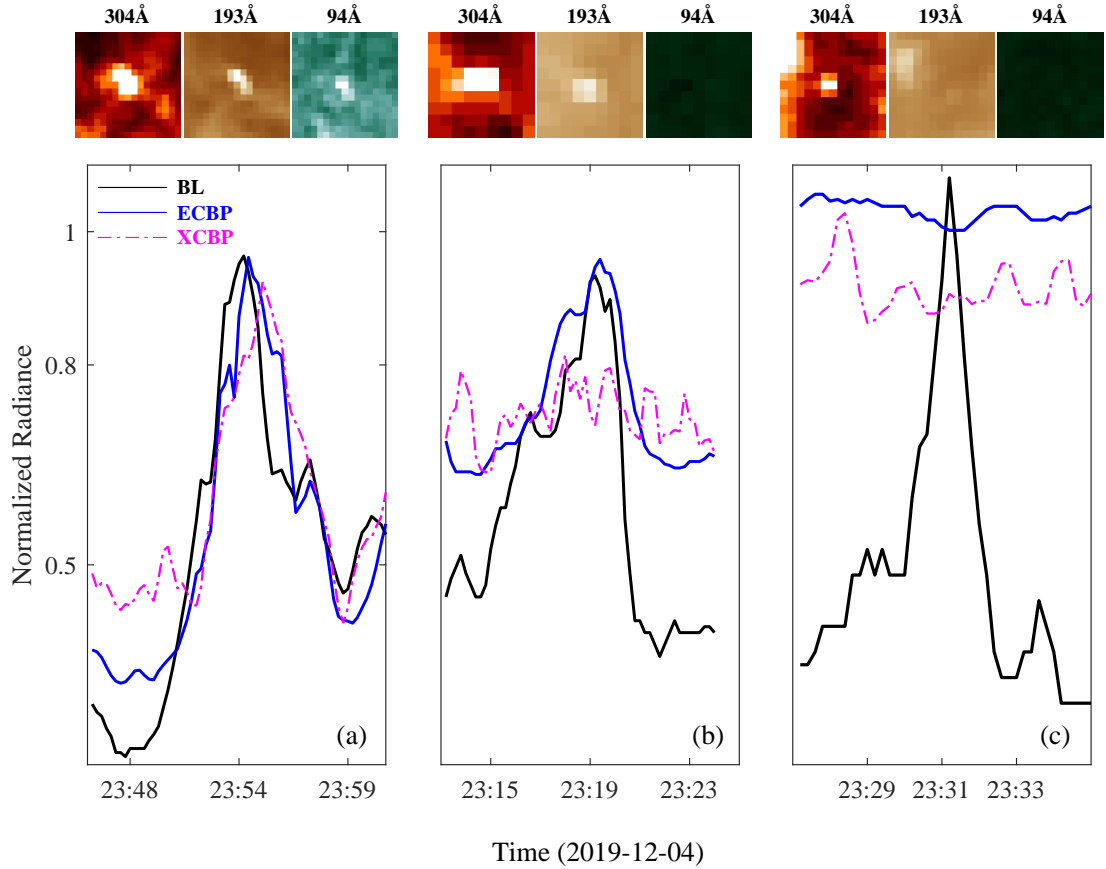}
 \caption{Lightcurves corresponding to blinkers in three filters 304, 193, and 94 \AA~ with time interval 12s. (a) Intensity enhancement in 304, 193, and 94 \AA~ simultaneously occurred blinker, ECBP, and XCBP at the position (x=-210{\arcsec} ,y=-123\arcsec). (b) Simultaneous intensity enhancement in 304 and 193 \AA~ occurred blinker and ECBP in the position (x=112{\arcsec} , y=179\arcsec). (c) Intensity enhancement in 304 \AA~ which occurred blinker in the position (x=-119{\arcsec} , y=80\arcsec). }
\label{fig3}
\end{figure}

\begin{figure}
\vspace{1cm}
\centering
\hspace*{-2.0cm}
\vspace*{-2.0cm}
\resizebox{6in}{!}{ \includegraphics{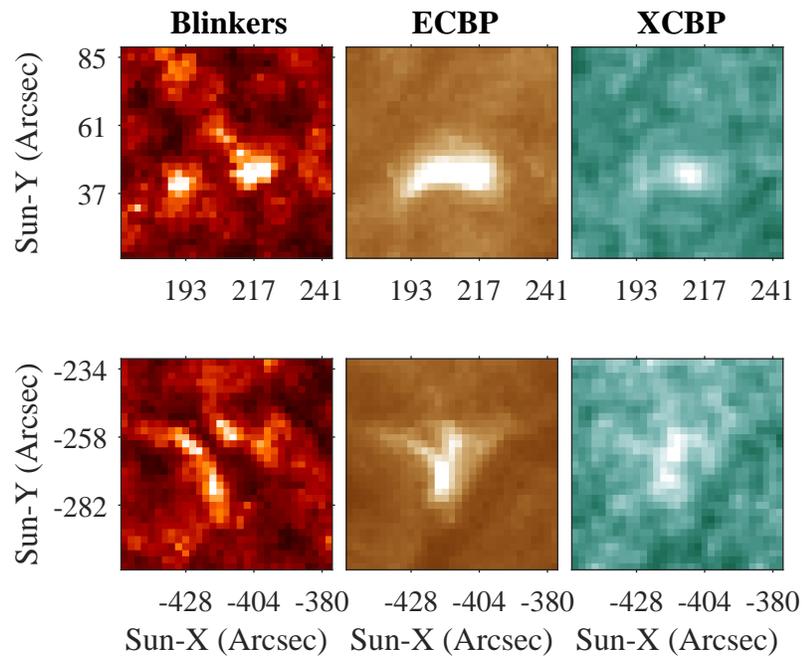} } 
 \caption{ Samples for the group of two (first row) and three (second row) blinkers associated with ECBP and XCBP.  }
\label{fig4}
\end{figure}

\begin{figure}
\vspace{1cm}
\centering
\hspace*{-1cm}
\resizebox{7.6in}{!}{ \includegraphics{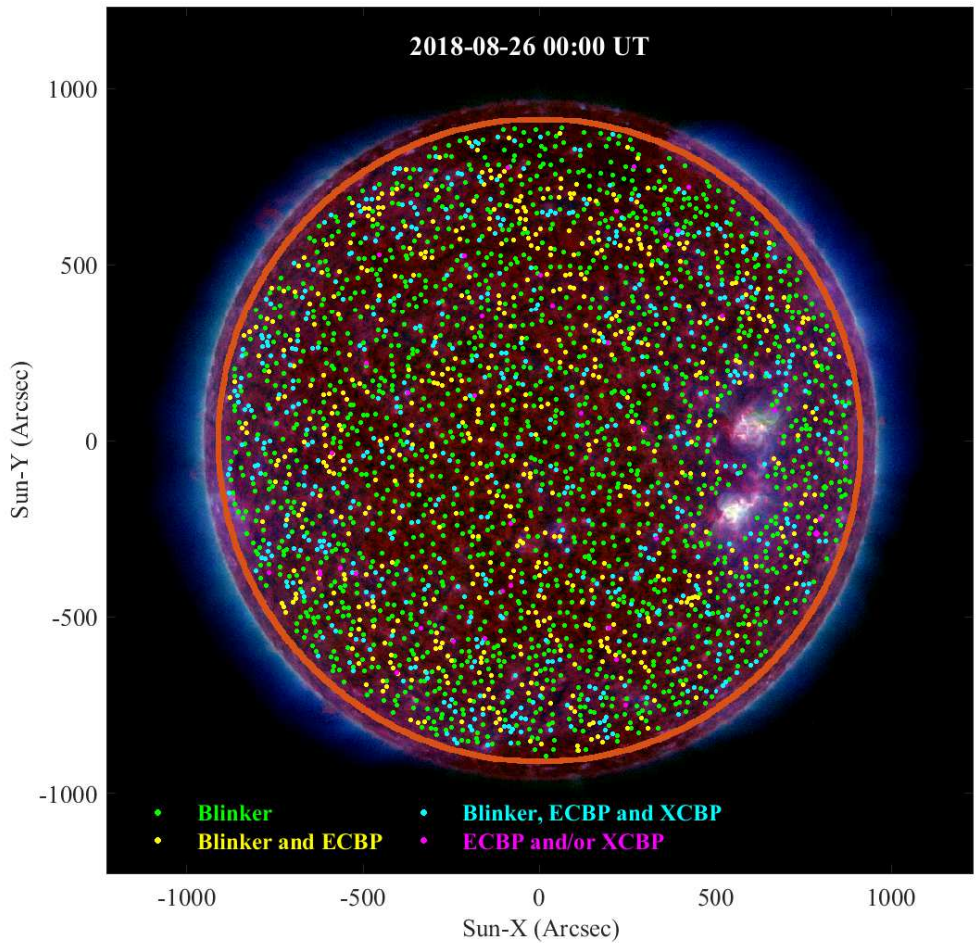} } 
 \caption{The SDO/AIA composite triple-filter image (304, 193, and 94 \AA) on 2018 August 26 00:00 UT. Location of QS blinkers (color points) recognized by presented automatic algorithm. The position of 46 ECBPs and/or XCBPs (purple points) detected without the related blinkers. To more avoid the high noisy features in the detection, we used the region inside the 0.95$R_\odot$ (red circle).  }
\label{fig5}
\end{figure}

In this work, first, we applied an automatic identification method to recognize the blinkers (at 304 \AA), ECBPs (at 193 \AA), XCBPs (at 94 \AA) from SDO/AIA images. Blinkers form on the transition region; however, ECBPs and XCBPs are the transient coronal features. Using a region-growing algorithm, we extracted pixels related to blinkers, ECBPs, and XCBPs. We considered an event (an ECBP and/or an XCBP) as the coronal counterpart of one or multiple blinkers if that blinker(s) appeared at a distance (boundary to boundary of features) less than 7.2{\arcsec} (empirically via trial and error) from the event. So, we studied the connection of blinkers, ECBPs, and XCBPs together. Figure \ref{fig3} shows lightcurves (with time interval 12 s) of transients at three AIA filters at 304, 193, and 94 \AA. We observed a blinker with the coronal counterpart features (ECBP and XCBP) is along with the lightcurves that were extracted with the simultaneous intensity enhancements at three filters versus the time (Figure \ref{fig3}(a)). Lightcurves indicate the simultaneous occurrence of a blinker and ECBP at 304 and 193 \AA, respectively (Figure \ref{fig3}(b)). Figure \ref{fig3}(c) represents a blinker without any coronal counterpart features. We also observed that groups (two or more) of blinkers are associated with a single coronal feature (Figure \ref{fig4}).    
Figure \ref{fig5} represents a composite triple-filter image (consisting of a 304, 193, and 94 \AA) with positions of blinkers observed on 2018 August 26 at 00:00 UT in 0.95$R_\odot$ (the area inside the red circle) at the solar disk. We found that out of 2785 blinkers (green points), about 1840 (yellow points) are associated with ECBPs, and about 978 blinkers (blue points) have occurred with both ECBPs and XCBPs. Furthermore, we observed 46 ECBPs (purple points) without detected blinkers. However, our investigation (by visual inspection) showed that the method did not detect 25 blinkers associated with ECBPs.     

Applying the automatic identification method, we observed 2198$\pm$478 (mean and standard deviation for daily number) QS blinkers for each $0.95R_\odot$ on the full-disk image. For the whole of the Sun's surface (about 2.2$\times$area$_{0.95\rm disk}$), we determined the total number of blinkers (sum of the number of blinkers at QS and ARs) at every moment. So, we estimated 5643$\pm$674 (average and one standard deviation) blinkers at every moment, which is twice the previous study \citep{harrison1997euv} at the \ion{O}{5}~ SoHO/CDS observations.
 \citet{brkovic2001analysis} followed a similar approach with an iterative scheme. They showed that the thresholds affect the birthrate. So, the number of blinkers increases to 20000. In this case, they showed that most blinkers were covered with a single-pixel (1.67\arcsec) or somewhat overlapping features. In the present method, we recognized events with individual pixels and diameters larger than 4{\arcsec} (to avoid the noise) without overlapping regions. Also, two or more features with a distance less than 4.8{\arcsec} are considered as a single event. We detect an average number of 612$\pm$152 ECBPs and 349$\pm$64 XCBPs for $0.95R_\odot$ images. So, we estimated 1346$\pm$334 and 767$\pm$141 of ECBPs and XCBPs for the whole of the Sun's surface, respectively. To compare the area (surface)  number density of QS and ARs blinkers, we first computed the QS area projected onto the Lambert cylindrical equal-area map. Then, we obtained the area number density of blinkers by dividing the QS and ARs blinkers number by the QS and ARs area, respectively. We obtained the surface density of blinkers about 0.001$\pm$0.0002 $\rm Mm^{-2}$ and 0.002$\pm$0.0005 $\rm Mm^{-2}$ for QS and ARs, respectively. These indicate that ARs' blinkers are more abundant than the QS \citep{parnell2002transition}.

\begin{figure}
\vspace{0.4cm}
\centering
\begin{center} 
\hspace*{-1.5cm}
\resizebox{9.45in}{!}{ \includegraphics{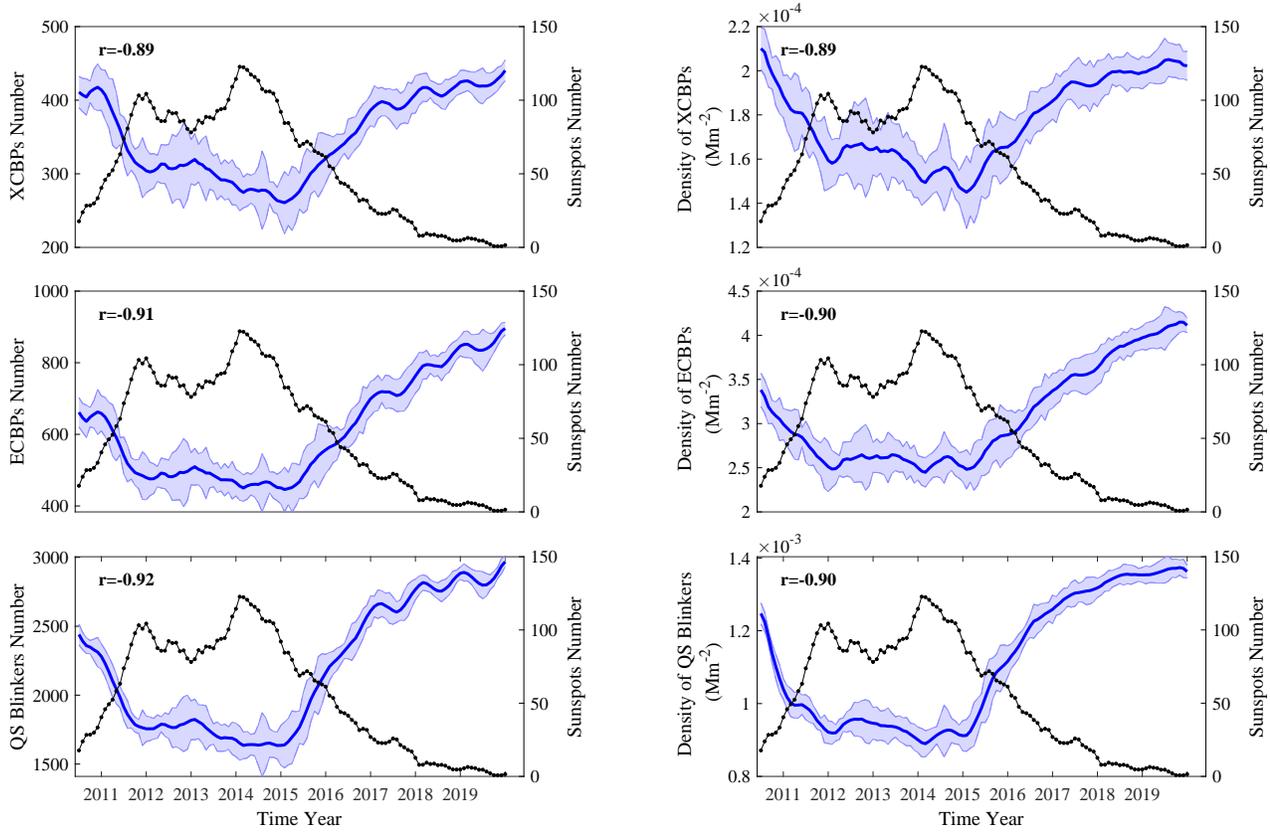} } 
\end{center} 
 \caption{The month-averaged time series of the sunspot numbers (dotted line), QS blinkers, ECBPs, and XCBPs (left panel), area number density of QS blinkers, ECBPs, and XCBPs (right panel) during ten years within solar cycle 24. The blue shaded area indicates the standard deviation of values (for a month) each time.}
\label{fig6}
\end{figure}

 Figure \ref{fig6} (left panel) shows the month-averaged time series of sunspot numbers (SILSO data, Royal Observatory of Belgium, Brussels), QS blinkers, ECBPs, and XCBPs during ten years within the solar cycle 24. We observed that transients and sunspots have very strong anti-correlation (Pearson correlations -0.93$<$ r $<$-0.88). In general, one may ask if the sunspot numbers and sizes increase during a cycle, the QS area decreases, which leads to the anti-correlation of QS transients with sunspots. We investigated the correlation between the area (surface) number density of transients and sunspots to address this question. Figure \ref{fig6} (right panel) represents the month-averaged time series of sunspot numbers, the area number density of QS blinkers, ECBPs, and XCBPs during the cycle. We observed that the area number density of QS transients has a very strong anti-correlation (Pearson correlations -0.92$<$ r $<$-0.88) with sunspots.

\begin{figure}
\hspace*{-7.5cm}
\vspace*{-1cm}
\centering
\resizebox{13in}{!}{\includegraphics{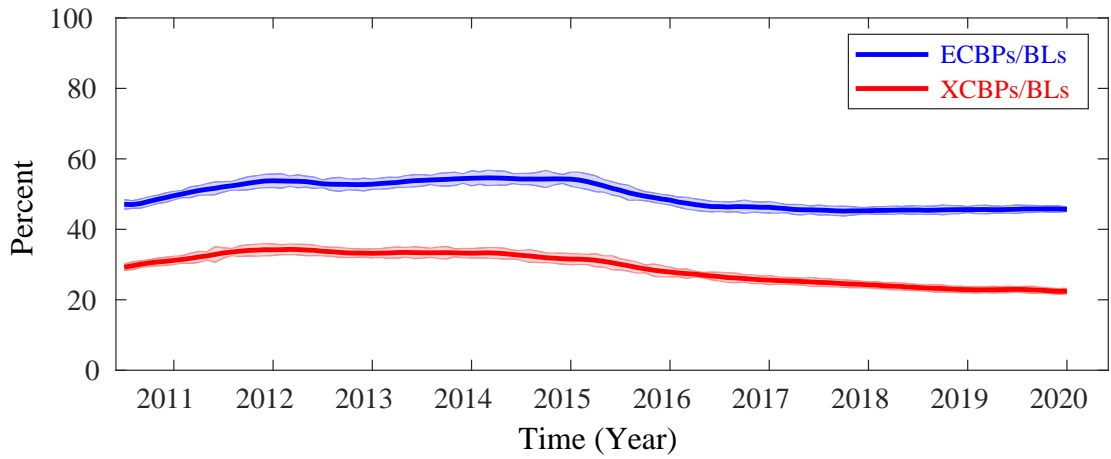}}
 \caption{The percent of blinkers are associated with ECBPs (blue line) and XCBPs (red line) during the solar cycle 24.}
\label{fig7}
\end{figure}

Using the ten years of AIA data with a cadence of 24 h, we identified in total 7,483,827 blinkers, 2,082,162 ECBPs, and 1,188,839 XCBPs. Our analysis shows that about 57\% and 34\% of blinkers are associated with ECBPs and XCBPs, respectively. Figure \ref{fig7} shows the percent of blinkers are associated with ECBPs (blue line) and XCBPs (red line) during the solar cycle 24. We observed that these percentages remain approximately constant during the cycle. The remaining blinkers with no coronal counterpart may be associated with chromospheric features, which are not investigated here. \cite{subramanian2012true} showed that two-third of blinkers are associated with the coronal counterpart features. These multiple connections between blinkers and counterpart features provide the link between the transition region and corona, and, hence, they play an important role in the mass and energy transfer in the solar atmosphere. Indeed, it seems the chromospheric upflows and coronal down-flows associated with blinkers may provide the emergence and preservation of the temperature gradient in the transition region. 

In order to obtain the events' birthrate, first, we tracked events within a rectangular box (with the size of 600$\arcsec\times$600$\arcsec$ with a cadence of 12 sec during 6 h (2019 September 00:00 UT- 06:00 UT). Then, we obtained the birthrate for events by dividing the number of events by the size of the box and the time interval (6 hours). We obtained the birthrate of $1.1\times10^{-18}$ ${\rm m}^{-2}{\rm s}^{-1}$, $3.8\times10^{-19}$ ${\rm m}^{-2}{\rm s}^{-1}$, and $1.5\times10^{-19}$ ${\rm m}^{-2}{\rm s}^{-1}$ for blinkers, ECBPs, and XCBPs, respectively. 

\subsection{Supergranular cell boundaries}
\begin{figure}
\vspace{2cm}
\centering
\includegraphics[width=\linewidth]{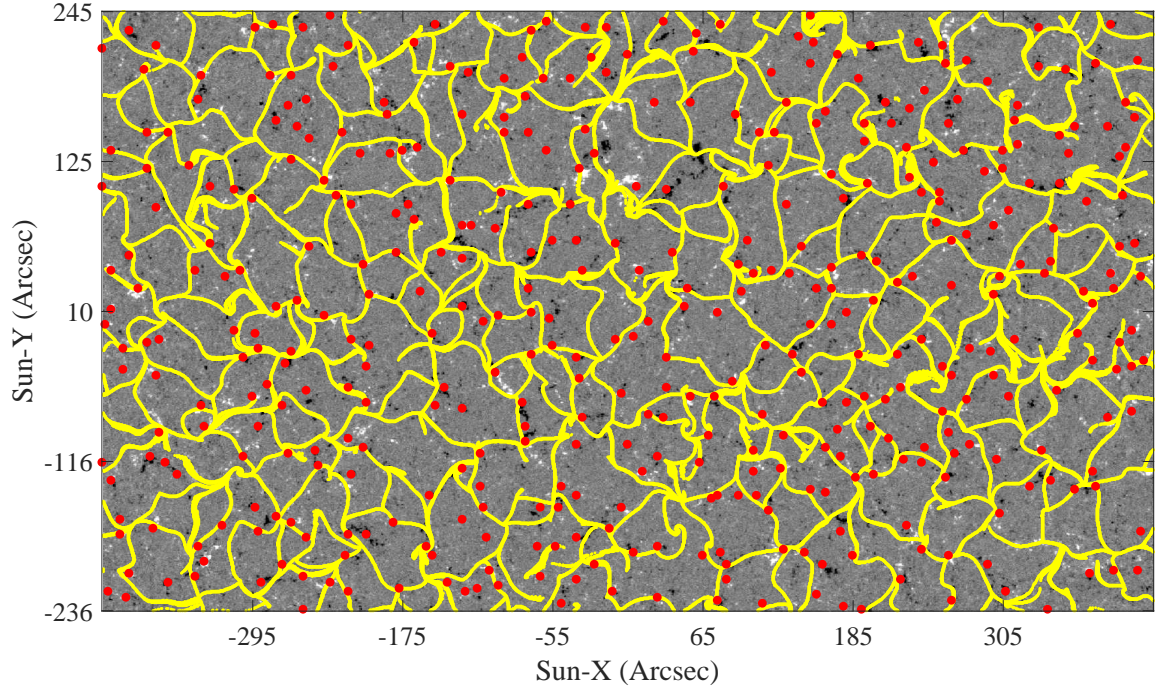}
 \caption{ SDO/HMI magnetogram with a field of view  840$\arcsec\times$480$\arcsec$. The location of 432 blinkers is marked with a red cycle. Superposed cell boundaries obtained by the ball-tracking method were mapped to an HMI magnetogram with a yellow line. Most blinkers are found close to supergranular boundaries.}
\label{fig8}
\end{figure}

It seems that blinkers may mostly occur at the boundaries of the supergranular cells. We applied a ball-tracking method based on the tracking solar photospheric flows on HMI continuum images to recognize the supergranular cell boundaries \citep{Potts2004,Attie2015, Attie2016}. We determined the supergranular cell boundaries  for a field of view 840$\arcsec\times$480$\arcsec$  on 2019 December 4 (00:00 to 00:30 UT). Figure \ref{fig8} shows the position of the peak intensity of 432 blinkers (red circles) and supergranular boundaries. By comparing the location of blinkers and supergranular cell boundaries, we found that about 80\% of blinkers appeared on boundaries and the remaining 20\% located within the cells, which is in agreement with \citet{yousefzadeh2016motion} in which about 90\% of the CBPs appeared on the supergranular cell boundaries. 

As blinkers have a short life and are pretty localized and considering that these features are analogous to those of flares and microflares (even if the amount of the involved energy is different), it might be possible to speculate that blinkers are generated through a mechanism similar to that which produces flares, i.e., magnetic reconnection \citep{Giovanelli_1946, Dungey_1953}, but at a different scale. With the help of high-resolution imaging and spectroscopic
observations, it is widely reported that the blinkers are the results of magnetic reconnection \citep{Priest_1999, marik2002, doyle2004, subr2008,young2018}. The blinker location at the supergranular cell boundaries might lead us to consider blinkers as events likely related to a reconnection mechanism occurring between adjacent magnetic structures with footpoints rooted at the supergranular boundaries. 
It generally makes sense to expect the more significant magnetic field variation among adjacent magnetic structures, making their border a favored candidate location for magnetic reconnection. 
Indeed, the horizontal flow dominates and sweeps the magnetic field towards the boundaries of supergranules. Hence, this process (horizontal flow) produces mixed polarity magnetic fields at the boundaries – the favorable topology of the magnetic field for magnetic reconnection. Hence, we can say that the boundaries of supergranules are the most favorable region for magnetic reconnection. 
Also, we showed that about 34\% of blinkers were related to the X-ray features (XCBPs). These may support categorizing some blinkers as small-scale flaring events. This speculation certainly needs to be supported by appropriate future observations. In addition, it is crucial to contemplate those reconnection mechanisms, such as tearing type instabilities \citep{Furth_et_al_1963} that have already been considered in the context of the solar corona \citep [see, e.g.,][]{Matthaeus_et_al_1984, Strauss1988, Velli_Hood_1989, Shibata2001, Dahlburg_et_al_2008, Pucci2014, DelSarto2016, Li2016, Singh_et_al_2019, Betar2020}, and three-dimensional reconnection models have been proposed \citep[see][]{Priest_1984, Janvier_2017}.

\subsection{Statistics of magnetic features}
\begin{figure}
\vspace{1.5cm}
\centering
\begin{center} 
\hspace*{-1.5cm}
\resizebox{6.45in}{!}{ \includegraphics{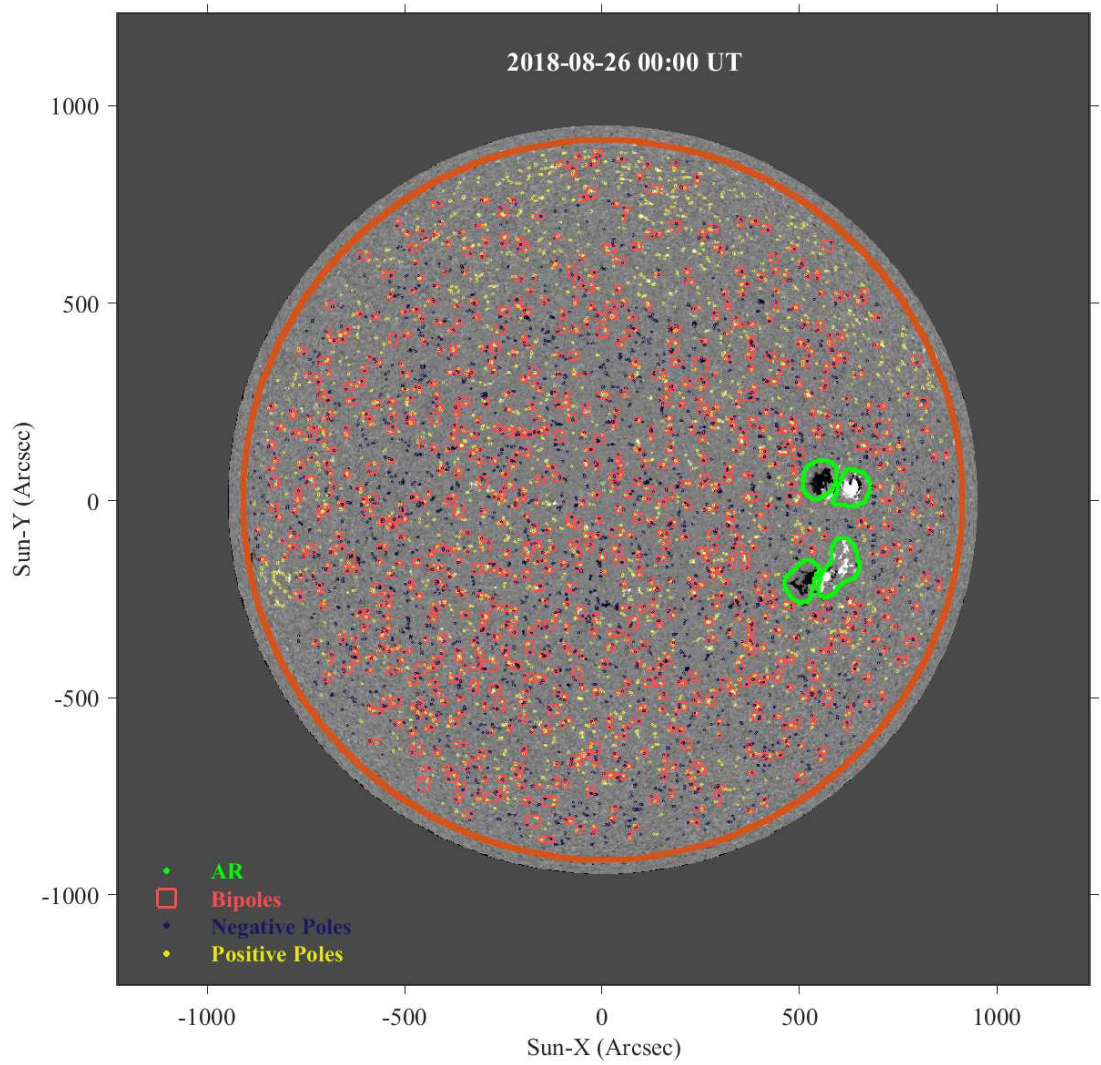} } 
\end{center} 
 \caption{The LOS magnetogram on 2018 August 26 00:00 UT including ARs boundaries (green points), positive poles (yellow points), negative poles (blue points), and bipoles (red boxes). }
\label{fig9}
\end{figure}

By applying the YAFTA code on the LOS HMI magnetograms, we extracted both QS and ARs small-scale magnetic poles with sizes in the range of  4.8{\arcsec} to 60{\arcsec} for the two thresholds ($>$ 12 G and $>$ 20 G). Figure \ref{fig9} shows ARs' boundaries (green contours) and all positive (yellow contours) and negative (blue contours) poles and bipoles (red boxes) in the LOS magnetogram (above 12 G). As we observed in the figure, out of 3890 poles, about 1842 of them are formed as bipoles. 

\begin{figure}
\vspace{2cm}
\centering
\begin{center} 
\resizebox{8.45in}{!}{ \includegraphics{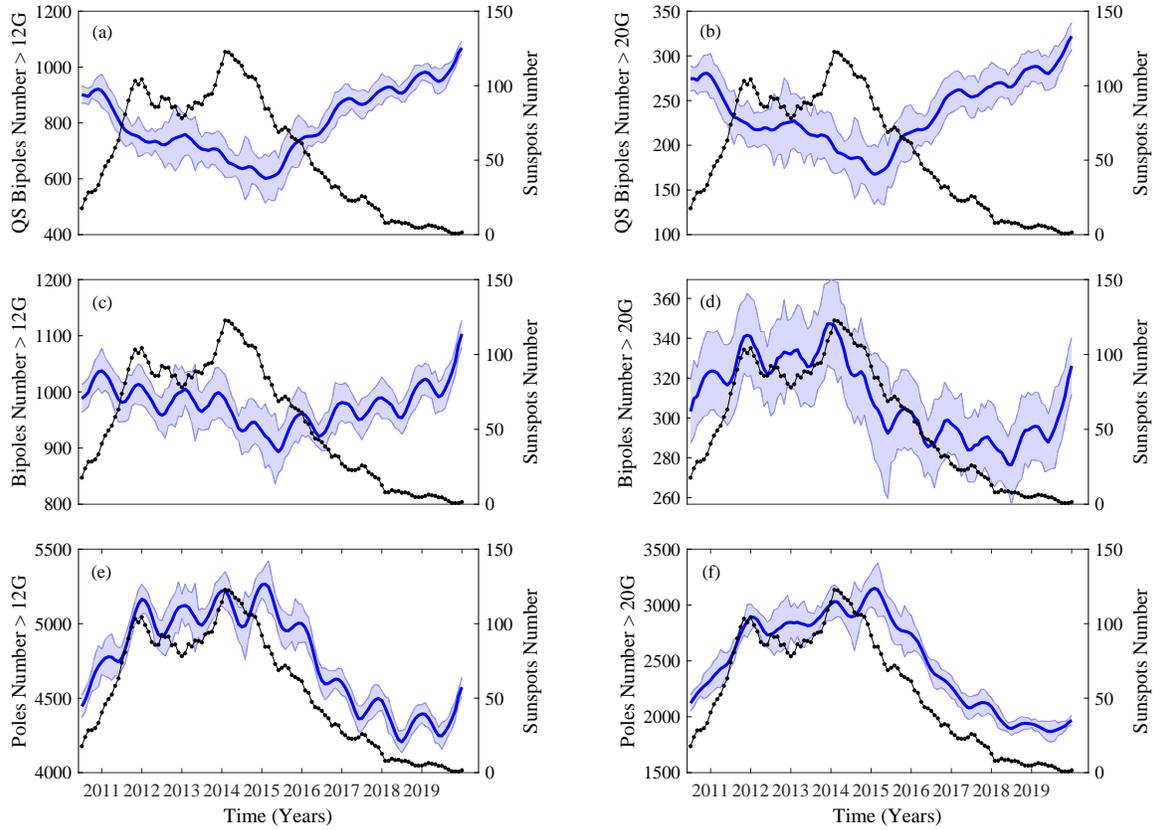} } 
\end{center} 
 \caption{The month-averaged time series of the sunspot numbers (dotted line), QS bipoles, bipoles (QS and ARs), and poles (QS and ARs) for threshold above 12 G (left panel) and 20 G (right panel) are shown. The blue shaded area indicates the standard deviation of values (for a month) each time.}
\label{fig10}
\end{figure}

Figure \ref{fig10}(a-b) depicts the month-averaged time series of sunspots and magnetic features (bipole and  poles). We found a strong anti-correlation (Pearson correlation r=-0.76 to -0.71) between the QS bipoles (for both thresholds) and sunspots. We also observed a strong anti-correlation (r=-0.78 to -0.72) between the QS poles and sunspots. We also obtained a strong anti-correlation between the area number density of QS small-scale magnetic features with sunspots, which confirms the anti-phased behavior. The number density was obtained by dividing the number of QS small-scale magnetic features to the QS area projected to the Lambert cylindrical equal-area map. The densities are still anti-correlated with sunspots, showing that this anti-phase is very likely not simply due to a decrease in the QS area during the solar maximum. This outcome might provide some insight into the solar dynamo mechanism.

We obtained very strong positive correlations (r$\ge$0.80) between the transients and QS bipoles. Previous studies \citep{Pre1999, Madjarsk2003, Ugarte-Urra2004} showed that most CBPs associated with QS bipoles. The negative correlation between the QS bipoles and sunspots conforms to the anti-cycle variation of transients \citep{Golub1979, Davis1983, Harvey1985}. The anti-phased behavior of the QS small-scale magnetic features and transients with sunspots and the locations of the transients at the supergranular boundaries are two important findings that verified the previous studies. Based on the identification of magnetic features of the whole solar cycle 23 from MDI/SoHO, \cite{Jin2011} and \cite{jin2012} obtained the anti-phase property for the QS  small-scale magnetic features with sunspots. Furthermore, \cite{Jin2011} stated that the small-scale magnetic features are generated by some local turbulent dynamo represented as an anti-phase with the global dynamo \citep[see also][]{SinKB_2018}.

The idea of a local and a global dynamo as two conceptually distinct dynamos has been proposed in the past years because of their diverse range of activity and because of the different properties in time, i.e., temporal coherence in the global dynamo, whereas time incoherence is in the local dynamo \citep{Proctor1995, stre1996, Cattaneo2001, Ossendrijver2003}. In recent years the idea that only one dynamo mechanism is operating in the Sun is finding its place in the community, and the point addressed in more recent works \citep{tobias2013, Pongkitiwanichakul_2016, Nigro2017} is how does one get a chaotic turbulent mess to have organization at large scale. Assuming this more recent scenario and the observed anti-correlation of the QS small-scale magnetic features with sunspots, the reason for this anti-correlation can be a self-organization process of the emerging magnetic flux into large-scale structures (i.e., ARs and hence sunspots) that should occur during the solar maxima, when magnetic energy increases. This large-scale self-organization might be due to the magnetic helicity inverse cascade by subtracting magnetic energy stored in the small-scale magnetic structures. 

Furthermore, we studied the correlation between the small-scale magnetic features of the full-disk and sunspots by adding the statistics of ARs' small-scale magnetic features to the QS for different thresholds.  
Figure \ref{fig10}(c) shows a very weak anti-correlation (r=-0.19) between bipoles (for the threshold $>$12 G) and sunspots. We observed a strong positive correlation (r=0.61) between bipoles and sunspots for the threshold $>$20 G (Figure \ref{fig10}(d)). 
We also found a strong positive correlation between poles (Figure \ref{fig10}(e-f)) and sunspots. Indeed, they both should provide their contribution to the large-scale components of the global magnetic field of the Sun (e.g., dipole and quadrupole component, assuming a decomposition of the global solar magnetic field on a spherical harmonic basis), thus poles and sunspots turn out to be in phase.

To study the synchronization of transients and their associated magnetic features, we considered 4.8{\arcsec} and 7.2{\arcsec} as the maximum distances (boundary to boundary) from magnetic poles to blinkers and coronal features, respectively. Out of 12,195,147 (5,694,525) detected poles, about 45\% (28\%) of them are the components of bipoles for a threshold above 12 G (20 G). We observed that 33\% (13\%), 55\% (26\%), and 59\% (32\%) of QS blinkers, ECBPs, and XCBPs are associated with QS bipoles for a threshold above 12 G (20 G), respectively. Also, we found that about 31\% (25\%), 26\% (28\%), and 25\% (30\%) of blinkers, ECBPs, and XCBPs, appear above regions where a unipole dominates at the threshold above 12 G (20 G), respectively. Overall, 61\%, 76\%, and 80\% of blinkers, ECBPs, and XCBPs appear above regions where one or both polarities at the threshold above 12 G and these percentages decrease to 37\%, 51\%, and 58\% with increasing the threshold to 20 G. \cite{bewsher2002transition} showed that about 88\% of blinkers (observed at \ion{O}{5}~ images) appeared above one pole or both polarities regions ($>$ 10 G and size $>$ 10 MDI pixels).   

Figure \ref{fig11} represents the scatter plot of the daily number of transients and the daily number of QS poles (with a threshold above 12 G). We observed that the daily number of blinkers, ECBPs, and XCBPs correlated with poles via the power-law relation ($y\propto x^\alpha$, where $y$ and $x$ indicate the transient numbers and pole numbers, respectively) with exponents 1.57$\pm$0.03, 1.84$\pm$0.03, and 1.36$\pm$0.01, respectively. This relation is useful to estimate the number of transients as a function of the photospheric QS magnetic field.         

Figure \ref{fig12} shows the cumulative distribution function (CDF) for the normalized maximum intensities (maximum intensity divided to the quiet background intensity) of QS transients and related magnetic flux of poles, as well as their scatter plots. To avoid the projection effects, we selected the events and their related magnetic flux within a central region with the size of 960$\arcsec\times$960$\arcsec$. We observed that the tail of the distributions follows the power-law behaviors (Figure \ref{fig12} left and middle panels). To obtain the power-law index ($\alpha$), we applied a maximum-likelihood fitting for data without binning \citep{Clauset2009}. This power-law behavior in the distribution tails can be due to a self-similar feature in the reconnection process that generates blinkers, analogous to what has been observed for flares and nanoflares as the system of self-organised criticality \citep{Lin1984, crosby1993, krucker1998, parnell2000, klimchuk2009, farhang2018, farhang2019}.
We also observed that (Figure \ref{fig12} right panel) the normalized maximum intensity of transients are correlated with the magnetic flux (the linear fits in log-log scale with slopes $\beta$ about 1.31, 0.94, and 0.97 (red lines) in the scatter plots) for blinkers, ECBPs, and XCBPs, respectively. To test the consistency for the power laws and correlations (slope of scatter plots), suppose $x$ and $y$ satisfy the power-law-like distributions as $N(x)\propto x^{-\alpha_x}$ and $N(y)\propto y^{-\alpha_y}$, respectively. Then, considering $y\propto x^\beta$ and after some algebra manipulation \citep[e.g.,][]{Aschwanden2011, Javaherian2017}, we give the slope as $\beta=(\alpha_x-1)/(\alpha_y-1)$. we obtained the slope about 1.49, 0.84, and 0.91 for the relation between the magnetic flux and the normalized maximum intensities for blinkers, ECBPs, and XCBPs, respectively.

\section{conclusion }\label{sec:summery}
Transients (such as blinkers, ECBPs, and XCBPs) are  the fundamental atmospheric phenomena that are mostly connected. These synchronized atmospheric features may transfer mass and energy to the different layers. Also, the chromospheric up-flows and coronal down-flows associated with blinkers may provide the emergence and preservation of the temperature gradient in the transition region. 

\begin{figure}
%\hspace*{-2cm}
\centering
\includegraphics[width=\linewidth]{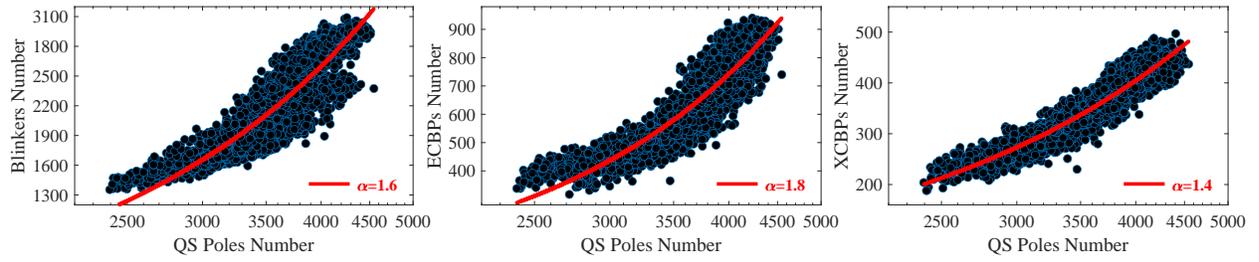}
 \caption{ Scatter plot of the daily number of QS poles (above 12 G) and the daily number of blinkers, ECBPs, XCBPs. We fitted a power-law curve $y\propto x^\alpha$ (red line) for each data.  }
\label{fig11}
\end{figure}

\begin{figure}
\vspace{2cm}
\hspace*{-1.5cm}
\centering
\resizebox{8in}{!}{\includegraphics[width=\linewidth]{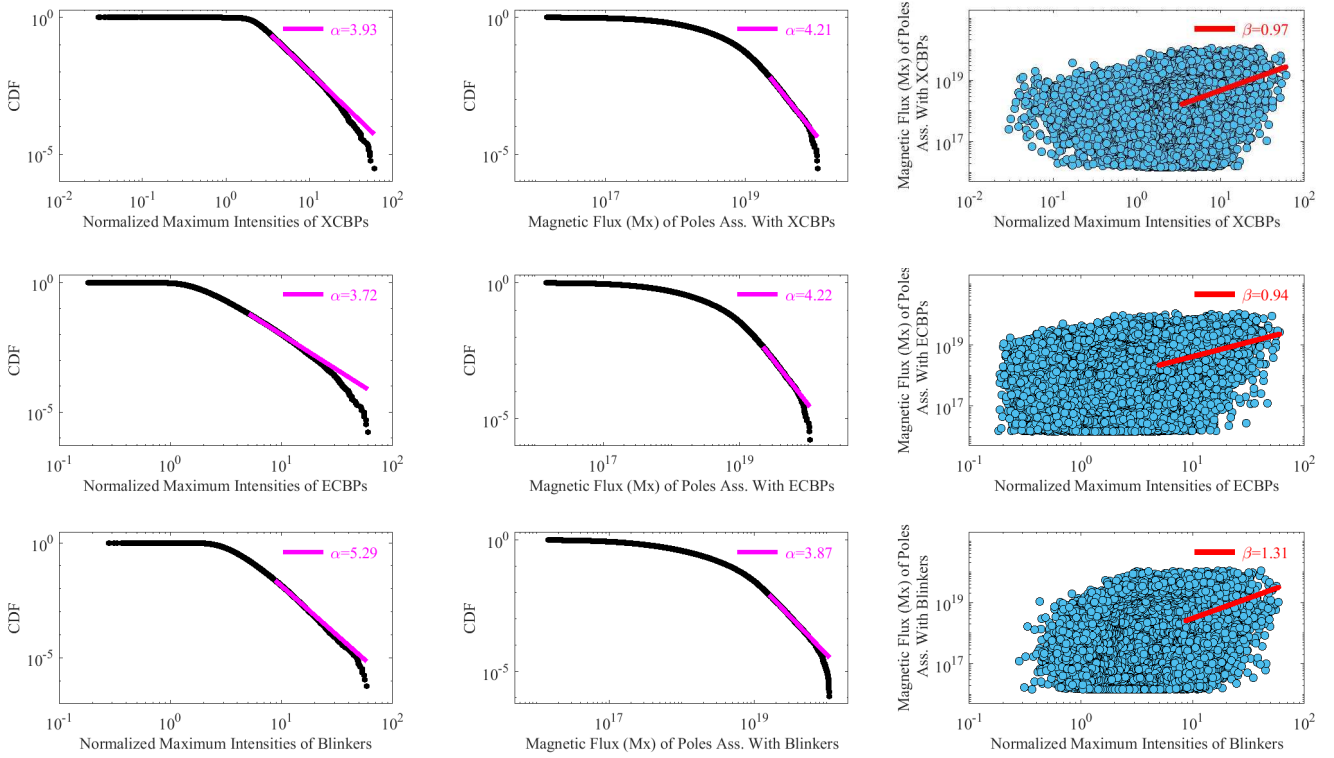} }
 \caption{\textbf{(Left panel) The cumulative distribution function (CDF) of the normalized intensities of transients (maximum intensity divided to the quiet background intensity), (Middle panel), CDF for the magnetic flux (in Mx unit) of poles associated with transients for the threshold above 12 G, (Right panels) Scatter plot for the magnetic flux of poles and normalized maximum intensity of transients. We fitted the power-law relation (purple lines) to the tails of distributions in which $\alpha$ indicates the power-law index. We fitted a linear function (red line) for each scatter plot (in log-log scale) at the regions corresponding to the tail of the distribution of the normalized maximum intensity.  }}  
\label{fig12}
\end{figure}

Here, we adopted an automated identification procedure based on ZMs and SVM classifier \citep{alipour2015statistical} to detect blinkers, ECBPs, and XCBPs at 304, 193, and 94 \AA~ AIA images, respectively. Classifiers contain features with various backgrounds, maximum intensities, shapes, morphologies, structures, and scales. The TSS was obtained greater than 0.86 as the performance of classifiers. Also, we applied the YAFTA algorithm on HMI magnetograms to extract the magnetic features such as poles and bipoles. We provide a summary of the main results  as follow:

\begin{itemize}
 \item[-]  We obtained an average number of 2198$\pm$478 blinkers for each image (in a 0.95$R_\odot$ of the solar surface) for ten years at solar cycle 24. Therefore, we estimated about 5643$\pm$674 blinkers for the Sun at every moment with a birthrate of $1.1\times10^{-18}$ ${\rm m}^{-2}{\rm s}^{-1}$.
 
  \item[-] By analyzing ten years of AIA data, we identified in total 2,082,162 (1,188,839) ECBPs (XCBPs). We also determined an average number of 1364$\pm$334  (767$\pm$141) at every moment for the entire of the Sun with the birthrate $3.8\times10^{-19}$ ${\rm m}^{-2}{\rm s}^{-1}$ ($1.5\times10^{-19}$ ${\rm m}^{-2}{\rm s}^{-1}$) for ECBPs (XCBPs).
   
   \item[-] Our investigation shows that roughly 57\% and 34\% of 7,483,827 blinkers at transition regions are associated with ECBPs and XCBPs, respectively. Therefore, these characteristics suggest that the transients are synchronized together at different layers of the solar atmosphere. In other words, the transients at different solar upper atmospheric layers tie up together probably with a similar physical process.

  \item[-] We showed that the majority of the blinkers occur near the supergranular cell boundaries, i.e., most of the blinkers are occurring in magnetic reconnection-prone regions.  Hence, this observational finding indicates (indirectly) that blinkers are possibly forming due to magnetic reconnection.

  \item[-] We observed a very strong anti-correlation between the number of sunspots, transients, and QS bipoles during solar cycle 24.  Assuming this anti-correlation is not a visible effect due to the background intensity variations, it can be due to an organization of the magnetic flux into large-scale structures as a consequence of a magnetic helicity inverse cascade that takes place when the magnetic energy increases during solar maxima. Indeed, when the magnetic energy increases via the omega effect, the magnetic energy is more efficiently channeled towards large scales, thus forming coronal loops and active regions. Consequently, the small-scale structures might have less energy and/or be in an inferior number, thus producing this observed anti-correlation.

\item[-] This study showed that roughly 61\%, 76\%, and 80\% of blinkers, ECBPs, and XCBPs are associated with QS poles at the photosphere, respectively.

 \item[-] The anti-phased behavior of the QS magnetic bipoles and transients with sunspots and the locations of the transients at supergranular boundaries confirm the local dynamo as an important mechanism for the evolution of quiet photosphere magnetic fields.
 
  \item[-]  We observed the power-law relation ($y\propto x^\alpha$) with $\alpha$ equals $1.57\pm0.03$, $1.84\pm0.03$, and $1.36\pm0.01$ for the daily number of blinkers, ECBPs, and XCBPs with the daily number of QS poles, respectively (Figure \ref{fig11}).    
  
 \item[-] We showed that the normalized maximum intensities of QS transients correlated with the magnetic fluxes via the power-law functions. The slopes of the power-law functions are verified by the power-law behavior for the normalized maximum intensities and  magnetic flux of poles distributions (Figure \ref{fig12}). The power-law behavior observed in the normalized maximum intensity distribution of transients leads us to consider these events as energy releases due to a self-similar process in analogy with what has been already observed for flares.  
 
\end{itemize}
  
In the next step, we would like to investigate the detection method for the small brightenings such as campfires from AIA/SDO and EUI on Solar Orbiter \citep{Chitta2021,Berghmans2021} to determine the role of the small-scale events in the heating of the coronal plasma.

\clearpage

\clearpage

\clearpage

\clearpage
\clearpage

\section*{Acknowledgement}
We thank NASA/SDO, HMI \& AIA science teams, for providing data used here. We also gratefully thank the anonymous referee for very useful comments and suggestions that improved the manuscript.

\clearpage
 \bibliographystyle{apj}
 \bibliography{ms}
 
\end{document}